\documentclass[conference]{IEEEtran}
\IEEEoverridecommandlockouts
\usepackage{cite}

\usepackage{amsmath,amssymb,amsfonts}
\usepackage{algorithmic}
\usepackage{graphicx}
\usepackage{subcaption}
\usepackage{multirow}
\usepackage{array}
\usepackage{colortbl}
\usepackage{tcolorbox}
\usepackage{textcomp}
\usepackage{pifont}
\usepackage{verbatim}
\usepackage{float}
\usepackage{tabularx}
\usepackage{placeins}
\usepackage{url}

\usepackage{tikz}
\usetikzlibrary{arrows.meta,positioning}
\usepackage[table,dvipsnames]{xcolor}

\def\BibTeX{{\rm B\kern-.05em{\sc i\kern-.025em b}\kern-.08em
    T\kern-.1667em\lower.7ex\hbox{E}\kern-.125emX}}

\begin{document}

\title{Escaping Local Minima: A Finite-Time Markov Chain Analysis of Constant-Temperature Simulated Annealing}

 \author{
 \IEEEauthorblockN{Hansini Ramachandran}
 \IEEEauthorblockA{\textit{Department of Electrical and Computer Engineering} \\
 \textit{University of Southern California}\\
 Los Angeles, California \\
 hramacha@usc.edu\vspace{-1.2em}}

 \and
 \IEEEauthorblockN{Bhaskar Krishnamachari}
 \IEEEauthorblockA{\textit{Department of Electrical and Computer Engineering} \\
 \textit{University of Southern California}\\
 Los Angeles, California \\
 bkrishna@usc.edu\vspace{-1.2em}}

}
\maketitle

\IEEEpubid{\parbox{\textwidth}{\scriptsize
\vspace{7\baselineskip}
© 2026 IEEE. Personal use of this material is permitted. Permission from IEEE must be obtained for all other uses, in any current or future media, including reprinting/republishing this material for advertising or promotional purposes, creating new collective works, for resale or redistribution to servers or lists, or reuse of any copyrighted component of this work in other works.
}}
\IEEEpubidadjcol

\begin{abstract}
Simulated Annealing (SA) is a widely used stochastic optimization algorithm, yet much of its theoretical understanding is limited to asymptotic convergence guarantees or general spectral bounds. In this paper, we develop a finite-time analytical framework for constant-temperature SA by studying a piecewise linear cost function that permits exact characterization. We model SA as a discrete-state Markov chain and first derive a closed-form expression for the expected time to escape a single linear basin in a one-dimensional landscape. We show that this expression also accurately predicts the behavior of continuous-state search up to a constant scaling factor, which we analyze empirically and explain via variance matching, demonstrating convergence to a factor of $\sqrt{3}$ in certain regimes.

We then extend the analysis to a two-basin landscape containing a local and a global optimum, obtaining exact expressions for the expected time to reach the global optimum starting from the local optimum, as a function of basin geometry, neighborhood radius, and temperature. Finally, we demonstrate how the predicted basin escape time can be used to guide the design of a simple two-temperature switching strategy. 
\end{abstract}

% Outline of the paper

% \begin{itemize}
%     \item We are presenting a closed-form mathematical analysis of the performance of Simulated annealing in a couple of idealized settings. 
%     \item First we consider a discrete 1-D function with a single linear basin. For this, we derive an expression for the expected time for a constant-temperature Simulated Annealing to leave the basin.
%     \item We show that this expression with an empirically determined constant factor can be used also to analyze the performance of a Simulated Annealing search on the continuous version of this 1-D function. We analyze this constant, and show how it tends to $\sqrt3$ in certain cases. 
%     \item We then extend our analysis to a 2-basin system, with a local optimum and a global optimum, allowing us to give an expression for the expected time to reach the global optimum. 
%     \item We use the analysis to demonstrate the existence of an optimum temperature (not too low or high) that minimizes the expected time to reach the global optimum.  
% \end{itemize}

\section{Introduction}

Simulated Annealing (SA) is a heuristic stochastic search algorithm that is widely used for many complex optimization problems ranging from Very Large Scale Integrated (VLSI) circuit placement and routing to quantum circuit compilation~\cite{sechen1985timberwolf,molavi2025qmr}. It mimics the physical process of bringing a system to its minimum energy state by controlling its temperature. 

While Simulated Annealing is empirically very successful, there are relatively fewer theoretical results characterizing its performance. Most prior results focus on steady state or asymptotic performance, or yield complex bounds that are not helpful for understanding finite state and time behavior.

To bridge this gap, this paper focuses on a simplified ``toy model" of the energy landscape. While real-world objective functions are high-dimensional and non-convex, we believe that studying simpler idealized settings could allow for the derivation of exact mathematical expressions that reveal the underlying dynamics of the search process. Unlike prior literature that often focuses on cooling schedules, emphasizing simplicity to gain insights, our work prioritizes the relationship between the geometric properties of the landscape and the expected hitting times under specific constant-temperature regimes.

Concretely, we present closed-form expressions for the behavior of constant-temperature SA in two discrete 1-D settings. First, we begin by analyzing a discrete 1-D piece-wise linear function characterized by a single basin, and derive an expression for the expected time required for a constant-temperature SA process to exit the basin. We then demonstrate the robustness and wider applicability of this expression by applying it to a continuous version of the same function, identifying an empirical constant factor that converges to $\sqrt{3}$ under specific conditions. %We further present an analysis for why it converges to this ratio.

Second, we analyze a dual-basin piecewise linear system that contains both a local and a global optimum. This extension allows us to formulate an expression for the expected time to reach the global optimum, providing a clear view of the dynamics of climbing out of a local optimum in the first basin to reach the global optimum in the second basin. %We examine the impact of various parameters such as the width to depth ratio of the basins, the neighborhood radius, and the temperature, on the time to reach the global optimum. 

%We also examine the empirical constant needed to approximate search over a continuous function in this case as well. One interesting finding is that for our simple setting, the performance (in terms of time to reach the global optimum) improves monotonically with an increase in the constant temperature. 

Third, we consider how the predicted first time to leave the suboptimal basin (obtained from our first analysis) could be used to determine an optimal time to switch from a high temperature to a low temperature. We show that this is indeed the case, finding empirically that the optimal switching time is generally monotonically increasing in the predicted mean time to leave the suboptimal basin (with an empirical quadratic fit offering a good regression).

These results serve as a foundational step toward a more rigorous, predictive theory of Simulated Annealing. By quantifying the interplay between temperature and landscape geometry in simple systems, we provide a framework that may eventually be scaled to higher-dimensional manifolds and gradual-temperature-reduction search processes, potentially guiding the development of more efficient self-tuning optimization algorithms. 

% \section{Related Work}
% %Here we identify key relevant papers on SA analysis and discuss what they show. Conclude with a para explaining how this paper is different from / complementary to those works. 

\section{Related Work}

The theoretical analysis of Simulated Annealing (SA) has progressed from asymptotic convergence guarantees to more practical finite-time characterizations. Our work lies in the latter category: by restricting to simplified landscape geometries, we derive exact closed-form expected escape and hitting times that complement existing spectral bounds and diffusion-limit perspectives.

Early theory established global convergence under logarithmic cooling \cite{geman1984stochastic}, and Hajek’s critical depth framework \cite{hajek1988cooling} clarified that schedules must effectively be slower than the escape time from the deepest local minimum. We adopt the notion of basin depth but focus on finite-time, constant-temperature regimes and quantify the expected time to escape.

For finite-time behavior, spectral methods bound failure probability via the Markov chain’s spectral gap \cite{desai1997finite,mitra1986convergence}, which is closely related to metastability results where escape times scale exponentially with barrier height \cite{holley1989asymptotics}. While these approaches yield rigorous general bounds, they often take the form of inequalities. By contrast, in idealized 1-D and two-basin toy landscapes—akin in spirit to piecewise-linear settings used to study mean first-passage times \cite{palyulin2012finite}—we obtain exact formulas that make the dependence on basin geometry (width and slope) and temperature explicit.

Our emphasis on constant temperature is also supported by prior work. Although adaptive strategies can offer speedups \cite{catoni1992rough}, Cohn and Fielding show that many landscapes admit an \emph{optimal fixed temperature} minimizing expected hitting time to the global optimum \cite{cohn1999simulated}. Finally, the $\sqrt{3}$ scaling factor we observe when relating discrete random walks to continuous processes connects to optimal-scaling theory for Random Walk Metropolis: Roberts, Gelman, and Gilks analyze diffusion limits and motivate variance matching between uniform proposals and the continuous limit \cite{roberts1997weak}, which explains the discretization mismatch and our radius-rescaling correction.

\section{Single-Basin Analysis}
\subsection{Assumptions}
Consider a one-dimensional, one-basin piecewise linear  function (with a decreasing slope from -N to 0, and with an increasing slope from 0 to N). We discretize the basin uniformly such that there are a total of 2N+1 states where $\pm N$ are the absorbing boundaries. 

\begin{figure}[h]
  \centering
  \includegraphics[width=0.7\linewidth]{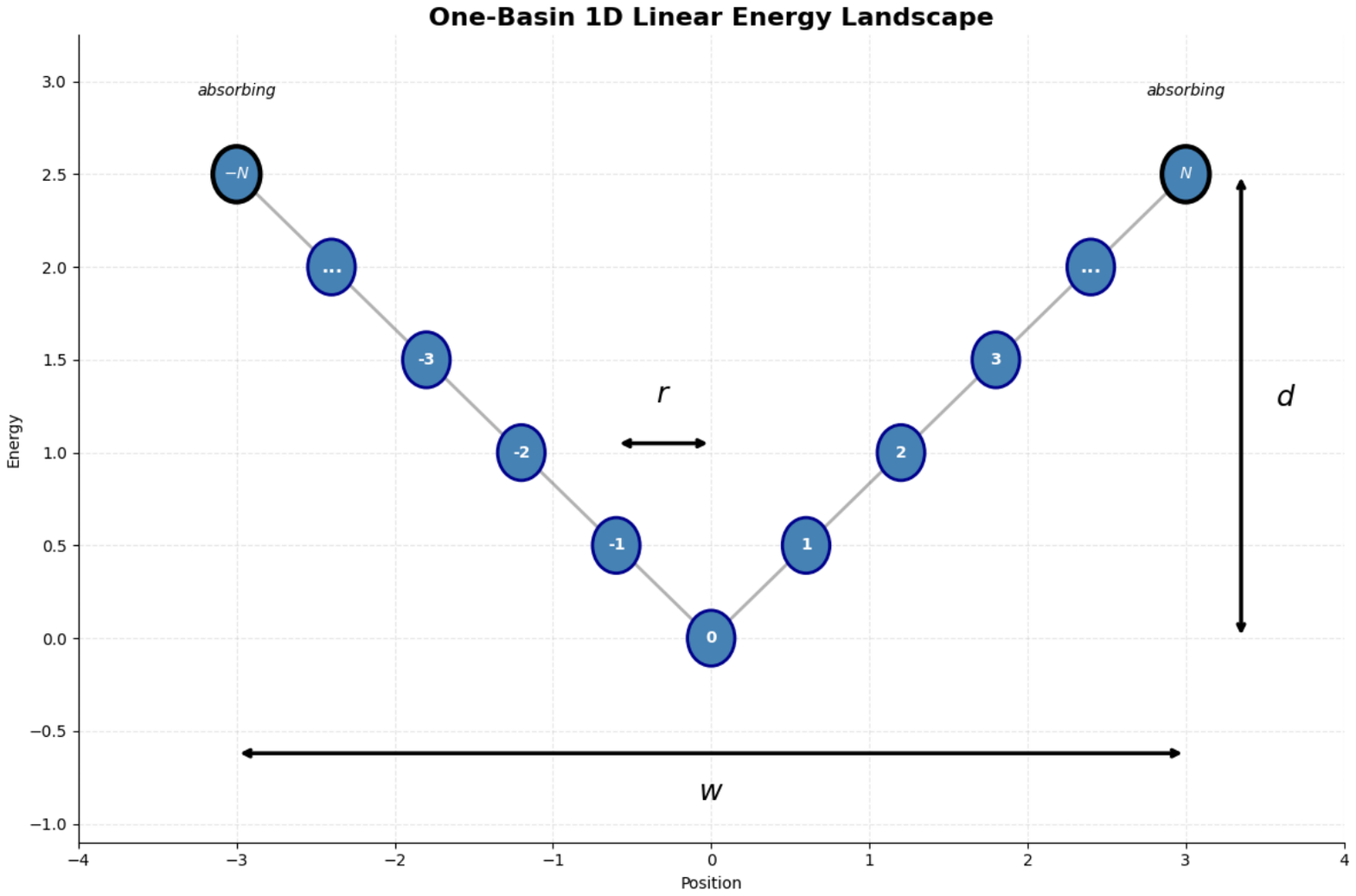}
  \caption{Energy landscape for a linear 1-basin geometry with physical parameters labeled}
  \label{fig:1D_optimalk_screenshot}
\end{figure}

For states \(i \in \{ 1, \ldots, N-1\}\):
\[
P_{i,i+1} = p, \qquad
P_{i,i-1} = 0.5, \qquad
P_{i,i} = 0.5 - p
\]

For states \(i \in \{-N+1, \ldots, -1\}\):
\[
P_{i,i-1} = p, \qquad
P_{i,i+1} = 0.5, \qquad
P_{i,i} = 0.5 - p
\]

For the center state \(i = 0\):
\[
P_{0,1} = P_{0,-1} = p, \qquad
P_{0,0} = 1 - 2p
\]

The boundary (absorbing) states satisfy:
\[
P_{-N,-N} = 1, \qquad P_{N,N} = 1 
\]
\subsection{Markov Chain / Random Walk Modeling}

\begin{figure}[h]
\centering
\begin{tikzpicture}[
  scale=0.7,
  transform shape,
  font=\small,
  >=Stealth,
  node distance=1.4cm,
  state/.style={circle,draw,minimum size=8mm,inner sep=0pt},
  absorb/.style={circle,draw,double,minimum size=8mm,inner sep=0pt},
  every loop/.style={looseness=8},
  thick
]

% --- Nodes ---
\node[absorb] (mN) {$-N$};
\node[state, right=0.7cm of mN] (m2) {$-2$};
\node[state, right=of m2] (m1) {$-1$};
\node[state, right=of m1] (z0) {$0$};
\node[state, right=of z0] (p1) {$1$};
\node[state, right=of p1] (p2) {$2$};
\node[absorb, right=0.7cm of p2] (pN) {$N$};

% --- Dotted connections to absorbing endpoints ---
\draw[densely dashed] (mN) -- (m2);
\draw[densely dashed] (p2) -- (pN);

% --- Self loops ---
\draw[->] (mN) edge[loop above] node[above=4pt] {$1$} (mN);
\draw[->] (m2) edge[loop above] node[above=4pt] {$\tfrac12 - p$} (m2);
\draw[->] (m1) edge[loop above] node[above=4pt] {$\tfrac12 - p$} (m1);
\draw[->] (z0) edge[loop above] node[above=4pt] {$1-2p$} (z0);
\draw[->] (p1) edge[loop above] node[above=4pt] {$\tfrac12 - p$} (p1);
\draw[->] (p2) edge[loop above] node[above=4pt] {$\tfrac12 - p$} (p2);
\draw[->] (pN) edge[loop above] node[above=4pt] {$1$} (pN);

% --- Left side transitions  ---

% -2 <-> -1
\draw[->] (m2) to[bend left=18]
  node[midway, above=6pt] {$\tfrac12$} (m1);
\draw[->] (m1) to[bend left=18]
  node[midway, below=6pt] {$p$} (m2);

% -1 <-> 0
\draw[->] (m1) to[bend left=18]
  node[midway, above=6pt] {$\tfrac12$} (z0);
\draw[->] (z0) to[bend left=18]
  node[midway, below=6pt] {$p$} (m1);

% --- Right side transitions ---

% 0 <-> 1
\draw[->] (z0) to[bend left=18]
  node[midway, above=7pt] {$p$} (p1);
\draw[->] (p1) to[bend left=18]
  node[midway, below=7pt] {$\tfrac12$} (z0);

% 1 <-> 2
\draw[->] (p1) to[bend left=18]
  node[midway, above=6pt] {$p$} (p2);
\draw[->] (p2) to[bend left=18]
  node[midway, below=6pt] {$\tfrac12$} (p1);

\end{tikzpicture}
\caption{Discrete Markov chain to model one linear basin with absorbing boundary states at $\pm N$.}
\label{fig:discrete_markov_absorbing}
\end{figure}

The governing second order difference expression for the above Markov Chain is:
\begin{equation}
(0.5 + p)T_{i} - 0.5T_{i-1} - pT_{i+1} = 1
\label{eq:MCdiff}
\end{equation}

The boundary condition is: 
\[
T_{0} = 1 +pT_{-1} + pT_{1} + (1-2p)T_{0}
\]

Applying symmetry, we get $
T_{0} - T_{1} = \frac{1}{2p} = \alpha $

To simplify the recurrence, define the first-difference sequence $
b_i \;:=\; T_{i}-T_{i+1},\qquad i\ge 1$.

Rewriting equation (\ref{eq:MCdiff}) as a first order difference equation:

\begin{equation}
b_i = \frac{1}{p} + \alpha b_{i-1}
\label{eq:bi}
\end{equation}

\subsection{Derived Expression for Expected Time to Exit}
Expressing equation (\ref{eq:bi}) as a recursive sum and telescoping up to the absorbing boundaries, 
we obtain a closed-form expression for the expected time required to reach
either absorbing state \(N\) or \(-N\), starting from an arbitrary initial
state \(i\). The resulting estimate for the mean absorption time is given
by
\begin{equation}
\begin{aligned}
T_i
&= \sum_{k=i+1}^{N} \alpha^k
 + \frac{1}{p} \Bigg[
   (N - i) \sum_{m=0}^{i-1} \alpha^m \\
&\qquad\qquad
 + \sum_{m=i}^{N-2} (N - m - i)\, \alpha^m
 \Bigg]
\end{aligned}
\label{eq:Ti}
\end{equation}

% \begin{equation}

% T_i = \sum_{k=i+1}^{N} \alpha^k
%   + \frac{1}{p} \Big[
%     (N - i) \sum_{m=0}^{i-1} \alpha^m
%     + (N - m - i) \sum_{m=i}^{N-2} \alpha^m \Big] \label{eq:Ti}  
    
% \end{equation}

where $\alpha = \frac{1}{2p}$.

The upward transition probability \(p\) may be expressed in terms of the
underlying physical energy landscape. For the one-dimensional basin with a constant-slope piece-wise linear walls,  where
\(w\) denotes the basin width, \(r\) the maximum radius of a proposed move,
\(d\) the basin depth, and \(T\) the temperature, the total number of discrete states is given by $N = \frac{w}{2r}$, and we can write $p = \frac{1}{2}\exp\!\Big(\frac{-2rd}{w\,\mathrm{T}}\Big)$.

%Substituting the
% corresponding expression for \(p\) yields

% \[
% \begin{aligned}
% T_i
% &=
% \sum_{k=i+1}^{\tfrac{w}{2r}}
% \exp\!\Big(\frac{2rdk}{wt}\Big)
% \\[6pt]
% &\quad
% +\,2\,\exp\!\Big(\frac{2rd}{wt}\Big)
% \Bigg[
% \Big(\frac{w}{2r}-i\Big)
% \sum_{m=0}^{i-1}
% \exp\!\Big(\frac{2rdm}{wt}\Big)
% \\[4pt]
% &\qquad\qquad
% +\,\sum_{m=i}^{\tfrac{w}{2r}-2}
% \Big(\frac{w}{2r}-m-1\Big)
% \exp\!\Big(\frac{2rdm}{wt}\Big)
% \Bigg]
% \end{aligned}
% \]

\subsection{Fit to Continuous Case}
The above formula for the estimated time to reach an absorbing boundary from a given interior state $i$ was derived for a discrete random walk Markov chain. We can use this to also model a  continuous Simulated Annealing random walk. 
\subsubsection{Empirical results}
In this section, we explore the fit between the estimated mean time predicted by the formula in eq. (\ref{eq:Ti}) for a discrete piece-wise linear 1-basin landscape versus the average number of steps needed to reach an absorbing boundary from state 0 (center) with a continuous SA. 

\begin{figure}[h]
  \centering
  \includegraphics[width=0.7\linewidth]{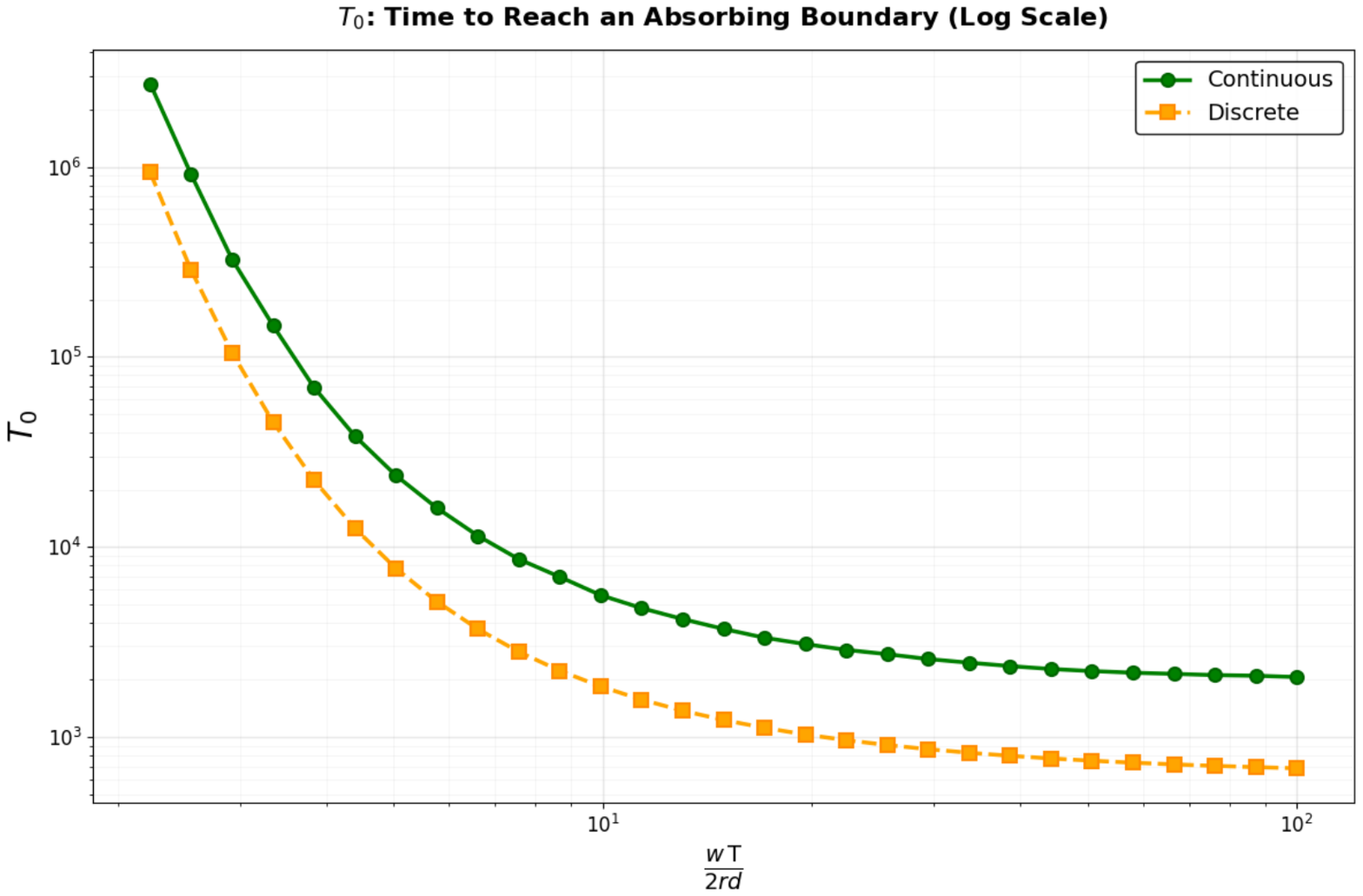}
  \caption{Estimated time to absorption (continuous and discrete) from state~0 vs.\ \(\tfrac{w\mathrm{T}}{2rd}\) ratio}
  \label{fig:1D_optimalk_screenshot}
\end{figure}

The dimensionless ratio \(\tfrac{w\mathrm{T}}{2rd}\) is varied from
approximately \(2.2\) to \(100\) by adjusting the basin depth in Fig.~3. As the depth decreases and the basin becomes shallower, \(\tfrac{w\mathrm{T}}{2rd}\) increases,
resulting in a reduced mean time to absorption. While both the continuous
Metropolis-averaged random walk and the discrete theoretical prediction
exhibit the same qualitative dependence on \(\frac{w\,\mathrm{T}}{2rd}\), their absolute
values do not coincide exactly. This systematic discrepancy highlights the effect of discretization on the predicted absorption times, despite the underlying agreement in scaling behavior.

The relative error between the two estimates of \(T_i\) remains approximately constant, with a flat error level of about \(67\%\) across the parameter range considered. This behavior indicates that the discrepancy arises from a multiplicative bias, suggesting that the predicted and observed absorption times are related by a constant scaling factor. Applying a corrective factor to the radius, rather than rescaling absorption time directly, is preferable because discretization alters the effective step variance and therefore addresses the root cause of the mismatch rather than its outcome.

\subsubsection{Explanation of $\sqrt{3}$ limit}

Because the discrete theoretical expression consistently overestimates the
observed mean absorption time, an empirical calibration was performed to
identify an optimal radius scaling factor that aligns the analytical
prediction with the Monte Carlo average computed over $2000$ independent
trials for identical physical parameters.

\begin{figure}[h]
  \centering
  \includegraphics[width=0.7\linewidth]{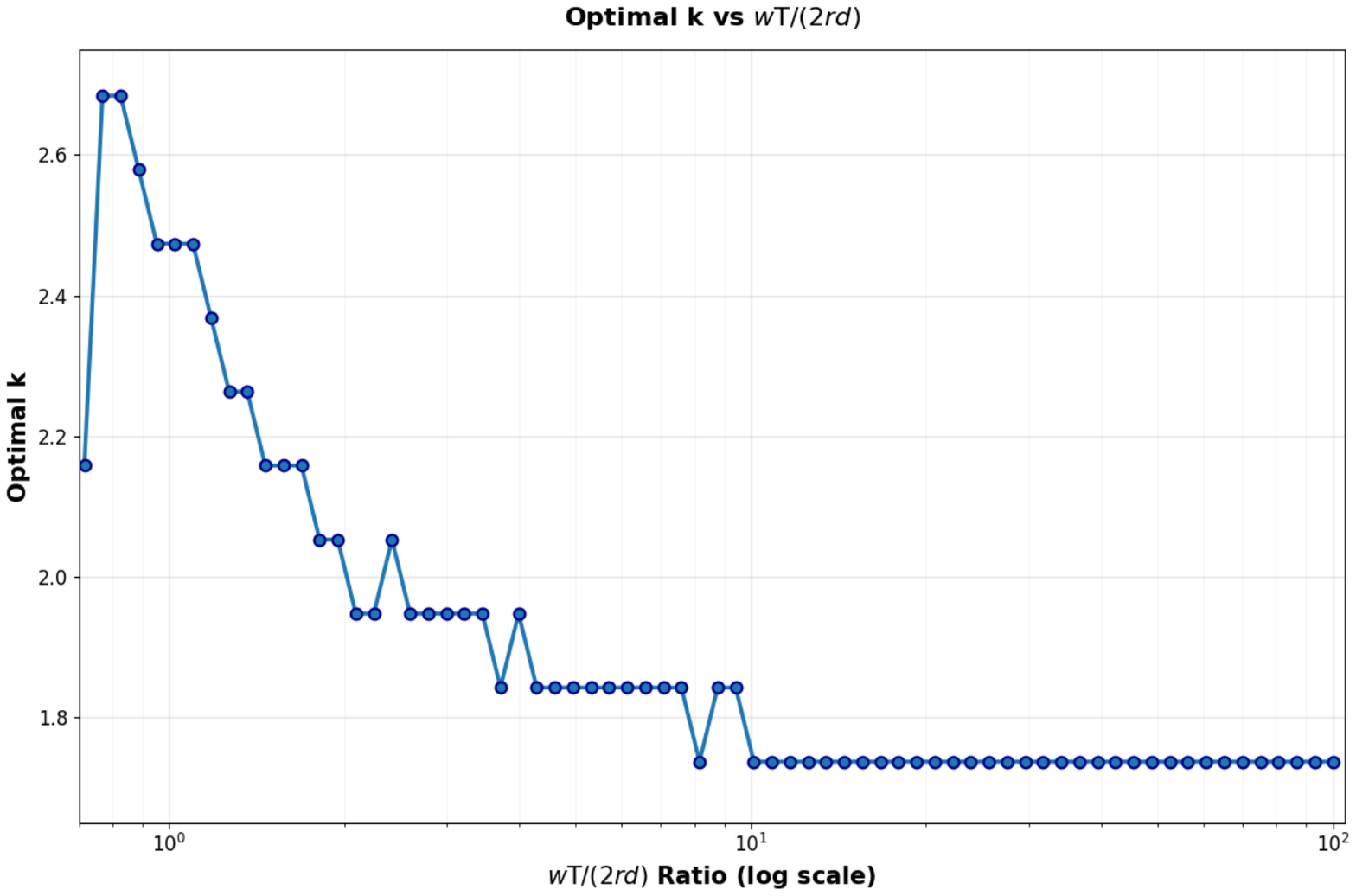}
  \caption{Optimal \(k\) (radius scaling factor) vs.\ \(\tfrac{w\mathrm{T}}{2rd}\) ratio}
  \label{fig:1D_optimalk_screenshot}
\end{figure}

As \(\tfrac{w\mathrm{T}}{2rd}\) increases, the ratio between \(r_{\mathrm{cont}}\) and
\(r_{\mathrm{disc}}\) converges asymptotically to \(\sqrt{3}\). This factor arises
from matching the effective drift and variance of the discrete random walk
to those of its continuous diffusion limit. In the continuous Metropolis
dynamics, the local motion can be approximated by a diffusion process whose
infinitesimal variance is proportional to the second moment of the proposal
distribution. In contrast, the discrete model uses a uniform proposal over a
finite interval of width \(2r_{\mathrm{disc}}\), whose variance is
\(r_{\mathrm{disc}}^2/3\). Matching this to the variance of the corresponding
continuous diffusion therefore requires scaling the discrete proposal radius
by a factor of \(\sqrt{3}\), yielding
\(r_{\mathrm{cont}} \approx \sqrt{3}\, r_{\mathrm{disc}}\). In the regime of large
\(\tfrac{w\mathrm{T}}{2rd}\), where boundary effects are negligible and the walk spends
most of its time in the interior of the basin, this variance matching
accurately captures the dominant contribution to the escape dynamics.
%This behavior is observed empirically in Figure~3.

After radius scaling, the relative error remains
below \(7\%\) in all but one of \(100\) trials, with a single outlier reaching
\(13\%\). This residual deviation arises from the stochastic nature of the
process, as each data point represents an average over \(2000\) independent
trials.

When \(\tfrac{w\mathrm{T}}{2rd}\) is small, the discrete walk overshoots the continuous
walk more severely. In this regime, finite-step effects and interactions with
the basin boundaries amplify the effective variance of the discrete dynamics,
leading to faster-than-predicted escape and necessitating a larger correction
factor \(k > \sqrt{3}\).

% Conversely, when $\tfrac{wT}{2rd}$ becomes very small
% (below approximately $0.7$), it is empirically observed that the scaling factor
% drops below $\sqrt{3}$ and approaches $1$. This occurs when the effective
% number of lattice states, $N = w/(2r)$, becomes small, causing the dynamics to
% be dominated by boundary effects rather than bulk diffusion, thereby
% invalidating the variance-matching approximation that underlies the
% $\sqrt{3}$ scaling.

% As $\tfrac{wT}{2rd}$ increases, the ratio between $r_{\mathrm{cont}}$ and
% $r_{\mathrm{disc}}$ converges asymptotically to $\sqrt{3}$. \textcolor{Red}{explain why root3.}This behavior is observed empirically in Figure 3. Figure 4 further shows that, after radius scaling, the relative error remains below $7\%$ in all but one of $100$ trials,
% with a single outlier reaching $13\%$. This deviation arises from the
% stochastic nature of the process, as each data point represents an
% average over $2000$ independent trials.

% When $\tfrac{wT}{2rd}$ is small, the discrete walk overshoots the
% continuous walk more severely. In this regime, boundary interactions
% and finite-step effects amplify the excess variance (and the corresponding
% escape rate), necessitating a larger correction factor $k > \sqrt{3}$. Further, when $\tfrac{wT}{2rd}$ is very small (below approximately $0.7$),
% it is empirically observed that the scaling factor drops below $\sqrt{3}$ and approaches ~1. This behavior arises when the effective number of lattice states,
% $N = w/(2r)$, becomes small, causing the dynamics to be dominated by
% boundary effects. 

With the proposal radius scaled by an optimal factor \(k\), the relative error
between \(T_{0,\mathrm{discrete}}\) and \(T_{0,\mathrm{continuous}}\) remains
consistently below \(9\%\). This level of agreement suggests that the discrete
Markov-chain--based expression provides a reasonable approximation to the
continuous mean first passage time \(T_0\) across the parameter regimes
considered. While the two models differ in their underlying state spaces, the
observed correspondence indicates that the discrete formulation captures the
primary dependence of the escape time on landscape geometry and temperature,
making it a useful analytical reference for continuous simulated annealing in
simplified settings.

\section{Two-Basin Analysis}
\subsection{Setup} 

In this section, we extend the single-basin framework to model a more general energy landscape containing multiple local minima and a single globally optimal minimum (although the formulation naturally generalizes to multiple optimal minima). The objective is to capture the dynamics of escape and convergence in the presence of competing basins, which more closely reflects realistic optimization landscapes. Unlike the single-basin case, the state space is now partitioned into two distinct basins separated by an energy barrier, and the effective slope within each basin need not be identical. This asymmetry introduces basin-dependent transition probabilities and escape rates, requiring separate characterization of the dynamics on either side of the barrier. While the underlying Markov structure and absorption framework remain unchanged, the two-basin model incorporates heterogeneous local geometry, enabling analysis of how basin depth and width jointly influence convergence behavior.

Consider a discrete Markov chain with two basins separated by a peak at \(M\)
and an absorbing state at \(N > M\).
The state space is extended symmetrically to the left and consists of
\[
-N, \ldots, -M-1, -M,\ldots, -1, 0, 1, \ldots, M, M+1, \ldots, N
\]
where \(T_i\) denotes the expected hitting time of state \(N\) when starting
from state \(i\).
The left boundary is reflecting, so that the dynamics for negative states are
symmetric to those on the positive side, and in particular
\(T_{-i} = T_i\) for all \(i \ge 1\).

\begin{figure}[H]
\centering
\begin{tikzpicture}[
  scale=0.6,
  transform shape,
  >=Stealth,
  thick,
  font=\footnotesize,
  node distance=1.0cm,
  state/.style={circle,draw,minimum size=7mm,inner sep=0pt},
  absorb/.style={circle,draw,double,minimum size=7mm,inner sep=0pt},
  dot/.style={inner sep=0pt,minimum size=0pt},
  every loop/.style={looseness=8}
]

% -------------------------
% Left omitted half (ellipsis)
% -------------------------
\node[dot] (dotsLeft) {$\cdots$};

% -------------------------
% Nodes
% -------------------------
\node[state, right=0.55cm of dotsLeft] (z0) {$0$};

\node[state, right=of z0] (one) {$1$};
\node[dot, right=0.55cm of one] (dotsL) {$\cdots$};
\node[state, right=of dotsL] (m1) {$M\!-\!1$};
\node[state, right=of m1] (m) {$M$};
\node[state, right=of m] (mp1) {$M\!+\!1$};
\node[dot, right=0.55cm of mp1] (dotsR) {$\cdots$};
\node[state, right=of dotsR] (Nm1) {$N\!-\!1$};
\node[absorb, right=0.6cm of Nm1] (N) {$N$};

% -------------------------
% Self-loops
% -------------------------
\draw[->] (one) edge[loop above] node[above=2pt] {$\tfrac12-p$} (one);
\draw[->] (m1)  edge[loop above] node[above=2pt] {$\tfrac12-p$} (m1);
\draw[->] (z0)  edge[loop above] node[above=2pt] {$1-2p$} (z0);
\draw[->] (mp1) edge[loop above] node[above=2pt] {$\tfrac12-q$} (mp1);
\draw[->] (Nm1) edge[loop above] node[above=2pt] {$\tfrac12-q$} (Nm1);

% -------------------------
% Left basin (near 0)
% -------------------------
\draw[->] (z0) to[bend left=18]
  node[midway, above=5pt] {$p$} (one);
\draw[->] (one) to[bend left=18]
  node[midway, below=5pt] {$\tfrac12$} (z0);

\draw[->] (dotsL) to[bend left=18]
  node[midway, above=5pt] {$p$} (m1);
\draw[->] (m1) to[bend left=18]
  node[midway, below=5pt] {$\tfrac12$} (dotsL);

% -------------------------
% Peak at M
% -------------------------
\draw[->] (m1) to[bend left=18]
  node[midway, above=5pt] {$p$} (m);
\draw[->] (m) to[bend left=18]
  node[midway, below=5pt] {$\tfrac12$} (m1);

\draw[->] (m) to[bend left=18]
  node[midway, above=5pt] {$\tfrac12$} (mp1);
\draw[->] (mp1) to[bend left=18]
  node[midway, below=5pt] {$q$} (m);

% -------------------------
% Right basin
% -------------------------
\draw[->] (mp1) to[bend left=18]
  node[midway, above=5pt] {$\tfrac12$} (dotsR);
\draw[->] (dotsR) to[bend left=18]
  node[midway, below=5pt] {$q$} (mp1);

\draw[->] (dotsR) to[bend left=18]
  node[midway, above=5pt] {$\tfrac12$} (Nm1);
\draw[->] (Nm1) to[bend left=18]
  node[midway, below=5pt] {$q$} (dotsR);

% Transition to absorbing state
\draw[->] (Nm1) to[bend left=18]
  node[midway, above=5pt] {$\tfrac12$} (N);

% -------------------------
% Absorbing state
% -------------------------
\draw[->] (N) edge[loop above] node[above=2pt] {$1$} (N);

\end{tikzpicture}
\caption{
Simplified representation of the two-basin Markov chain showing only the
right-hand half of the state space. The omitted left-hand side consists of a reflected copy with
identical transition structure. State \(M\) denotes the barrier between
basins, and \(N\) is an absorbing state.
}
\label{fig:two_basin_mc}
\end{figure}
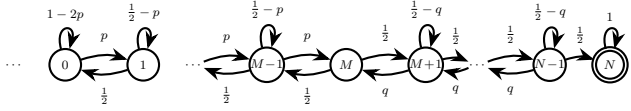

\subsection*{Left Basin (\(1 \le i \le M-1\))}
For states in the left basin, the transition probabilities are
\[
P(i \to i+1) = p, \qquad
P(i \to i-1) = \tfrac12, \qquad
P(i \to i) = \tfrac12 - p .
\]

where we take $p=\frac12 \exp(-2 r d_1/(w_1 T))$ for the left basin of width $w_1$ and depth $d_1$.

\subsection*{Peak at \(M\)}
At the barrier separating the two basins, transitions are unbiased:
\[
P(M \to M-1) = \tfrac12, \qquad
P(M \to M+1) = \tfrac12 .
\]

\subsection*{Right Basin (\(M+1 \le j \le N-1\))}
Within the right basin, the transition probabilities are
\[
P(j \to j-1) = q, \qquad
P(j \to j+1) = \tfrac12, \qquad
P(j \to j) = \tfrac12 - q .
\]

where we take $q=\frac12 \exp(-2 r d_2/(w_2 T))$ for a right basin of width $w_2$ and depth $d_2$.

\subsection*{Absorbing State}
The terminal state \(N\) is absorbing:
\[
T_N = 0, \qquad P(N \to N) = 1 .
\]

\subsection*{Symmetry at the local minimum (state 0)}

At the point of symmetry, at state 0, the transition probabilities satisfy
\[
P(0 \to 1) = p, \qquad
P(0 \to -1) = p, \qquad
P(0 \to 0) = 1 - 2p .
\]
By symmetry, \(T_{-i} = T_i\), so \(T_{-1} = T_1\).

The first-step equation for \(T_0\) is therefore
\[
T_0
= 1 + (1 - 2p) T_0 + p T_1 + p T_{-1}
\]

For the left basin (indices $0\le i\le M$), define
\begin{equation}
D_i \;:=\; T_i - T_{i+1},\qquad 0\le i\le M,
\end{equation}

Because the Markov chain is symmetric on both sides, we simplify and define the first difference as below with a boundary condition at state 0. 
\[
D_0 = T_0 - T_1 = \frac{1}{2p} \equiv \alpha .
\]
\[
D_i = \frac{1}{p} + \alpha D_{i-1}.
\]

To replace the left basin recursion at the peak (state M), 
\[
D_M = 2 + D_{M-1}
\]

For the right basin (indices $M\le j\le N-1$), define
\begin{equation}
B_j \;:=\; T_j - T_{j+1},\qquad M\le j\le N-1.
\end{equation}
We can write the difference equation for this basin as:
\[
B_j = 2 + 2q B_{j-1}.
\]
where q is the upward probability from states $M < j < N$
where the boundary condition for the landscape is 
\[
B_M = D_M.
\]

\subsection{Expression}
The first-difference equations governing the left and right basins were
solved using a common recurrence structure and a telescoping-sum
argument. The boundary condition at the peak state was then applied to
couple the two solutions, yielding a consistency relation between the
left- and right-basin difference sequences \(D_i\) and \(B_j\).
{\small
\[
T_i =
\left\{
\begin{array}{@{}l@{}}
\displaystyle
\sum_{k=i}^{M-1}
\left(\frac{1}{p} \sum_{j=0}^{k-1} \alpha^j + \alpha^{k+1}\right)
+ \sum_{s=0}^{R-1}
\left(2 \sum_{u=0}^{s-1} \beta^u + \beta^s D_M\right)
\hfill 0 \le i \le M,
\\[0.6em]
\displaystyle
\sum_{s=i-M}^{R-1}
\left(2 \sum_{u=0}^{s-1} \beta^u + \beta^s D_M\right)
\hfill M+1 \le i \le N-1.
\end{array}
\right.
\]
}

where
\[
D_M = 2 + \frac{1}{p} \sum_{j=0}^{M-2} \alpha^j + \alpha^{M}.
\]

and
\[
\alpha = \frac{1}{2p} ; 
\beta = 2q
\]

\subsection{Empirical fit for the Continuous Case}
To calibrate the discrete theoretical prediction against the averaged
continuous-time absorption statistics, the same radius-scaling procedure
was applied. In the two-basin setting, the correction factor was found to
depend primarily on the dimensionless ratio
\(
\frac{w\,\mathrm{T}}{2 r d}
\)
associated with the suboptimal basin, provided that the optimal basin is
sufficiently wide and deep
\(
\left( \frac{w}{d} \gtrsim 5 \right).
\)
In this regime, the dominant source of discretization-induced mismatch
arises from the suboptimal basin.

Here, the basin is sufficiently deep that once the walker
crosses \(M\), the probability of traversing back up the optimal basin
rather than moving toward the minimum is very small. Consequently, changes
in the proposal radius—affecting step size and local transition
probabilities—do not accumulate through repeated stochastic retries.
In other words, although the number of steps from \(M\) to \(N\) scales deterministically
with
\(
\frac{w}{2r},
\)
this dependence does not induce a multiplicative bias beyond the explicit
geometric scaling already captured in the discrete model. As the ratio
\(
\frac{w}{d}
\)
increases further, this already small multiplicative effect decreases
rapidly. Empirically, it is observed that as long as
\(
\frac{w_2}{d_2} \gtrsim 5
\)
(with the basin width also sufficiently large to yield a sufficient number of discrete states), the influence of the optimal basin on the fitted radius correction factor remains small. When this scaling factor \(k\) is applied, the relative error between the scaled discrete and continuous mean first passage times remains consistently below \(5\%\) over the corresponding range of \(
\frac{w_1\,\mathrm{T}}{2 r d_1}
\).

This suggests that, in this regime, the escape dynamics are primarily governed by the geometry of the suboptimal basin, and that the discrete formulation—after radius scaling—provides a reliable approximation to the continuous dynamics.

\begin{figure}[h]
  \centering
  \includegraphics[width=0.7\linewidth]{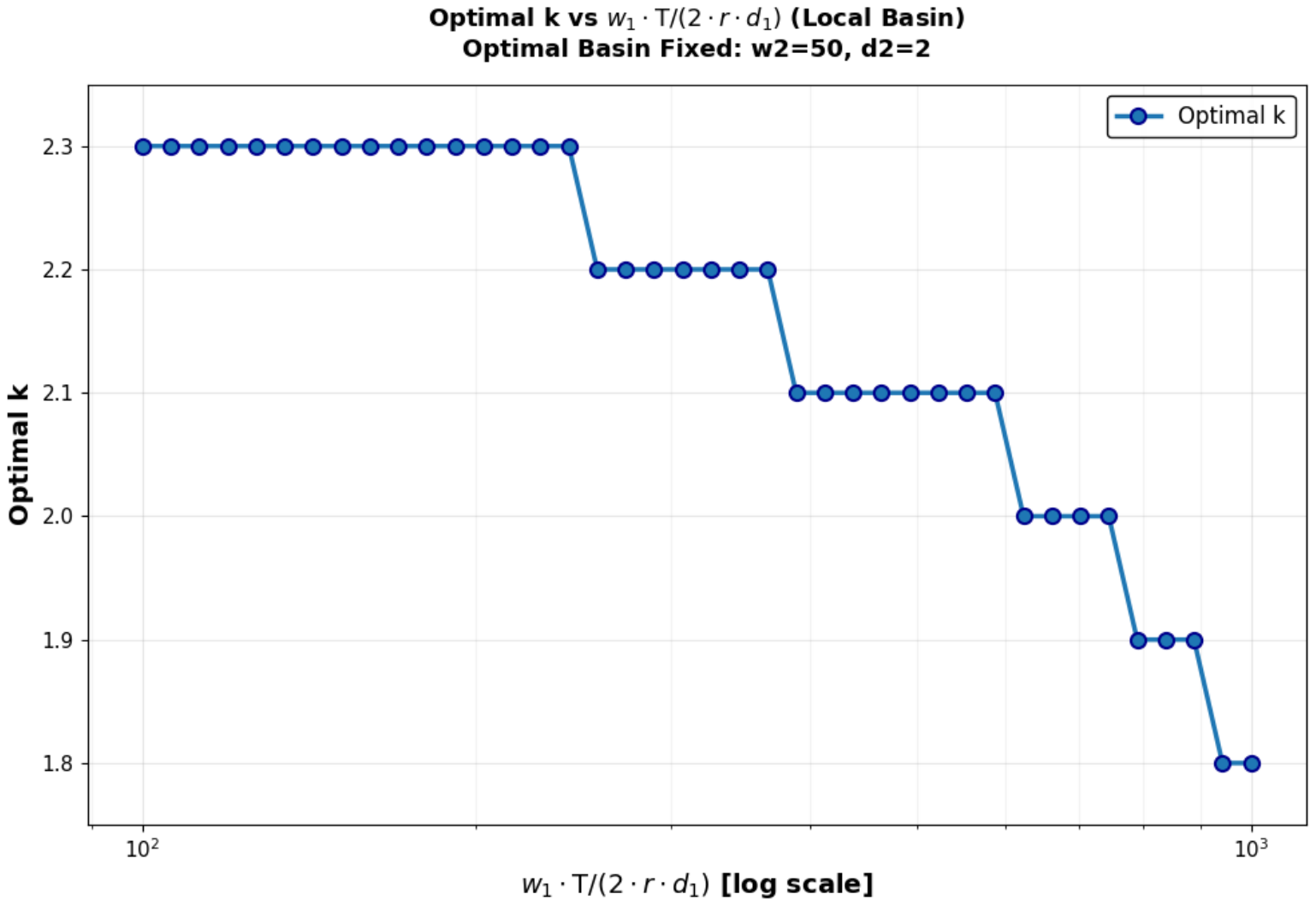}
  \caption{Optimal \(k\) as a function of the ratio \(\tfrac{w_1\,\mathrm{T}}{2 r d_1}\), with the optimal-basin geometry held fixed at \(w_2 = 50\) and \(d_2 = 10\).}
  \label{fig:2d_optimalk}
\end{figure}

\subsection{Exploration: Determining the Impact of Temperature}

\paragraph{Temperature}
We consider how the temperature impacts the hitting-time to reach the global optimum. Increasing $T$, which increases uphill acceptance (via larger $p$ and $q$), is always better (i.e., monotonically reduces the hitting time), as seen in Fig. ~\ref{fig:temperature_sweep_application}.

\begin{figure}[H]
  \centering
  \includegraphics[width=0.7\linewidth]{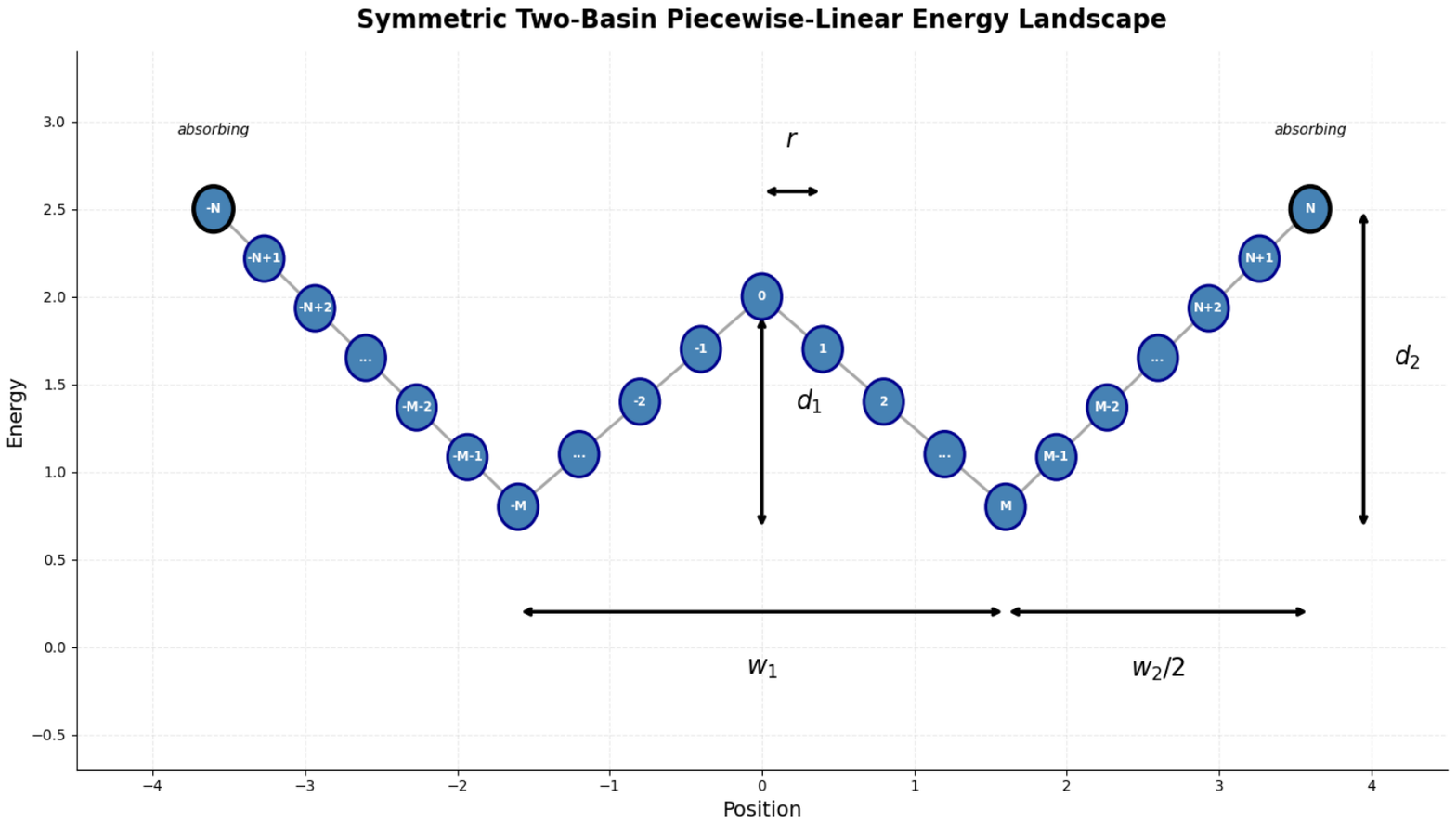}
  \caption{Energy landscape for a linear 2-basin geometry with physical parameters labeled}
  \label{fig:1D_optimalk_screenshot}
\end{figure}

The temperature is swept from \(\mathrm{T} = 1\) to \(\mathrm{T} = 50\) for several values of \(\tfrac{w}{d}\) (3.5, 3.75, 4.0, and 4.25), while keeping the proposal radius fixed
at \(r = 1\). Only the discrete formula is used in this analysis.

\begin{figure}[h]
  \centering
  \includegraphics[width=0.7\linewidth]{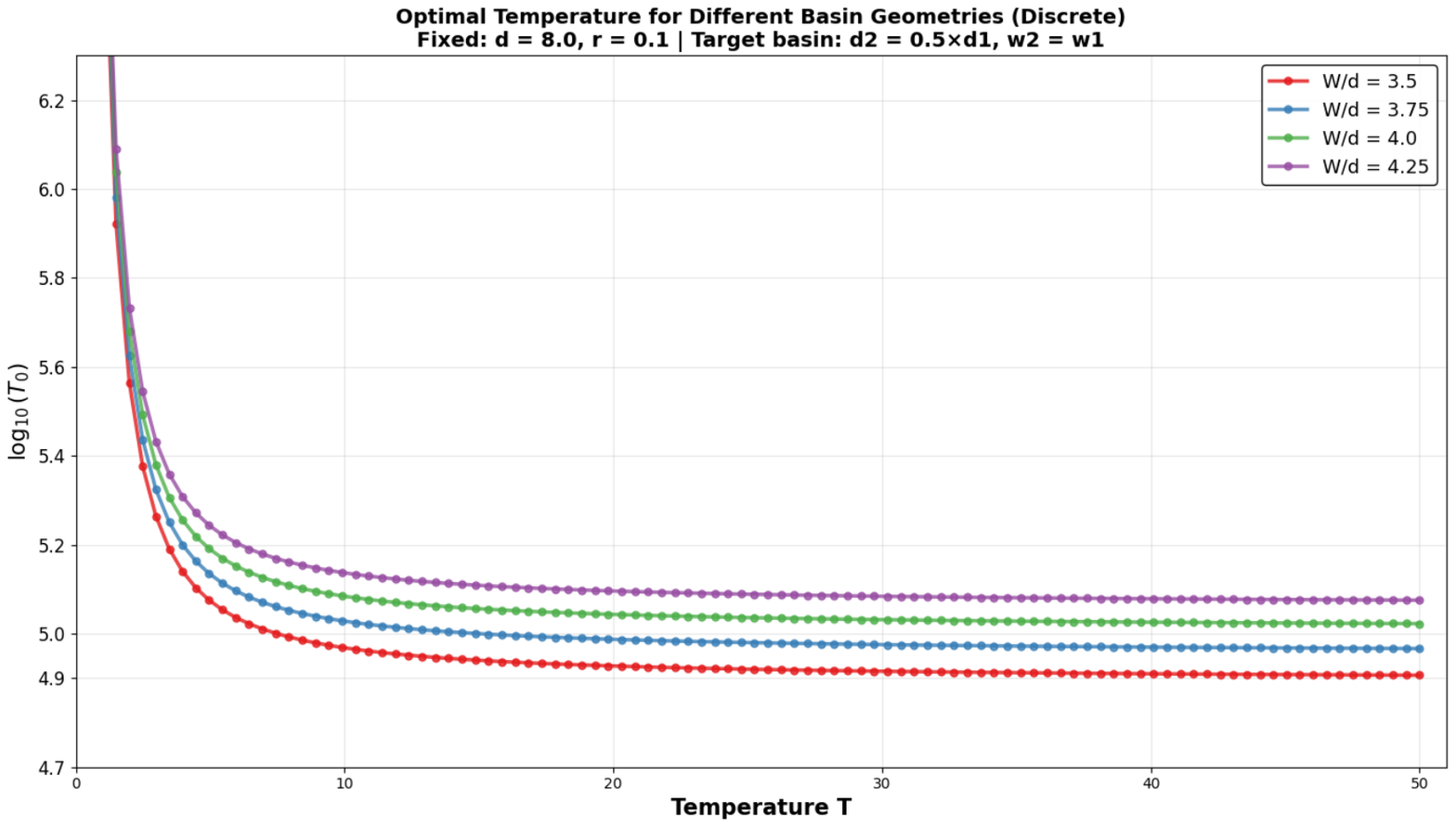}
  \caption{Plot of mean first passage time vs. temperature for different basin geometries $\frac{w}{d}$}
  \label{fig:temperature_sweep_application}
\end{figure}

\subsection{Finding an optimal time to switch temperature}

An optimal switching time for a two-basin geometry arises from a basic trade-off in constant-temperature simulated annealing dynamics. At a higher temperature, uphill moves are accepted more readily,
which increases the probability of escaping the suboptimal basin. However, maintaining a high temperature after the barrier has
been crossed can be counterproductive: the chain continues to accept energetically unfavorable moves, increasing backtracking and reducing the tendency to reach the global minimum. Switching to a lower temperature after escape therefore can reduce $T_0^{\mathrm{global}}$ by ensuring the walker moves primarily downhill.

To illustrate the practical value of the closed-form escape-time prediction, we
evaluate a simple two-temperature schedule that switches abruptly from a high
temperature to a low temperature. Specifically, we fix
\(\mathrm{T}_{\mathrm{high}}=10\) and \(\mathrm{T}_{\mathrm{low}}=3\), and we run SA at
\(\mathrm{T}_{\mathrm{high}}\) for \(\tau\) steps before switching to
\(\mathrm{T}_{\mathrm{low}}\)
for the remainder of the run. For each landscape configuration,
\(\tau\) is swept over a set of candidate values parameterized as
\(\tau = c\,\widehat{t}_{\mathrm{escape}}\), where
\(\widehat{t}_{\mathrm{escape}}\) is the predicted mean time to leave the local
basin obtained from the discrete expression derived in Section~III, and \(c\) is
a scalar multiplier.

The dashed lines in Fig.~9 correspond to the baseline mean first-passage time
\(T_{0}\) when the temperature is held fixed at \(\mathrm{T}_{\mathrm{high}}\)
throughout. In contrast, the two-temperature policy can reduce \(T_{0}\) for an
appropriate choice of \(\tau\), and the empirical results exhibit a clear
minimizer over the tested multipliers \(c\). This confirms that (i) a nontrivial
optimal switching time exists for this setting, and (ii) the predicted
local-basin escape time provides a useful scale for selecting the
switching-time search range. In practice, the closed-form prediction can
therefore be used to efficiently parameterize and narrow the exploration of
switching policies, reducing the amount of empirical tuning required.

\begin{figure}[h]
  \centering
  \includegraphics[width=0.7\linewidth]{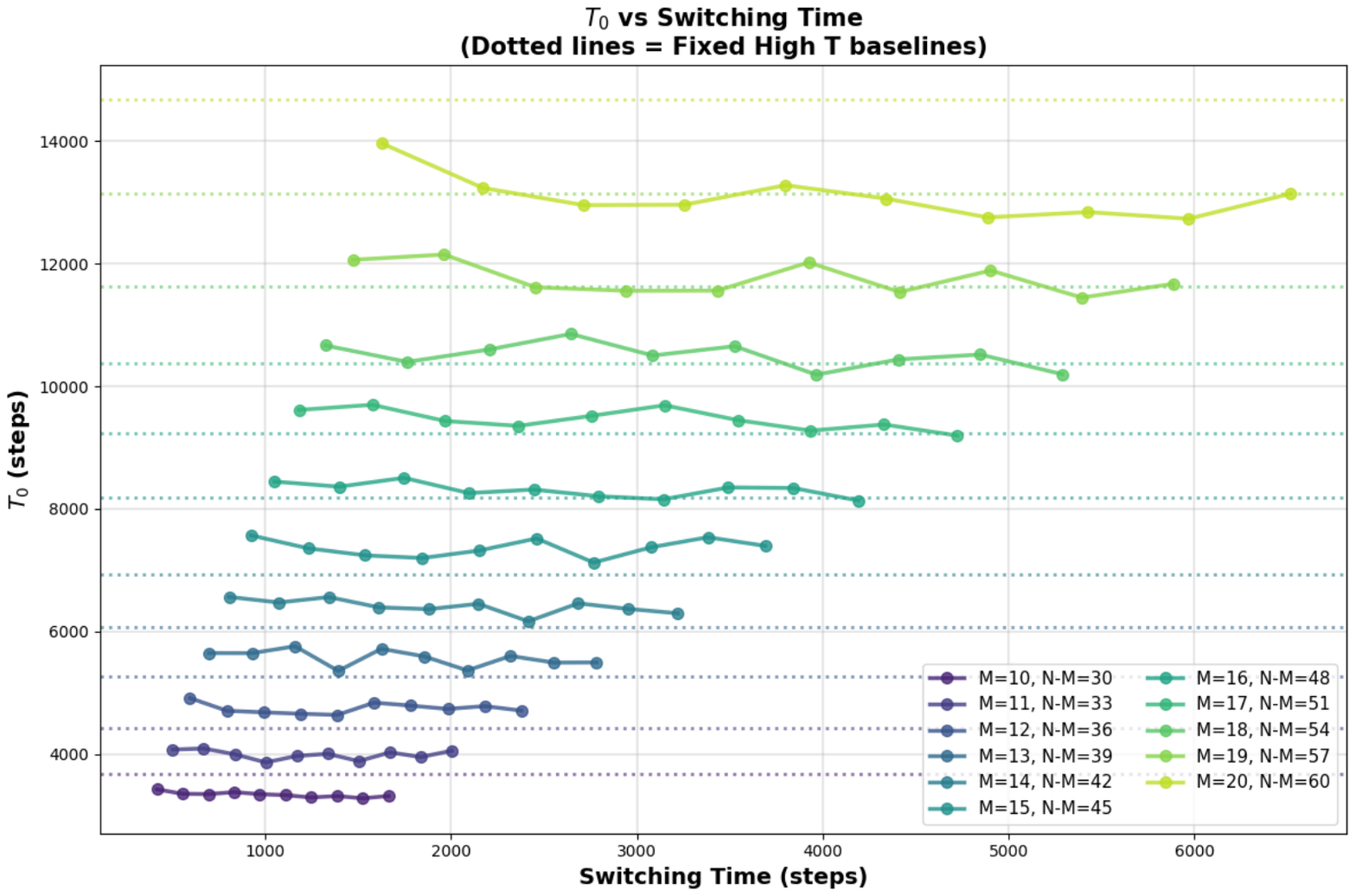}
  \caption{Plot of mean first passage time for 2 basin geometry versus predicted switching time for different M/N basin configurations}
  \label{fig:temperature_sweep_application}
\end{figure}

\begin{figure}[h]
  \centering
  \includegraphics[width=0.7\linewidth]{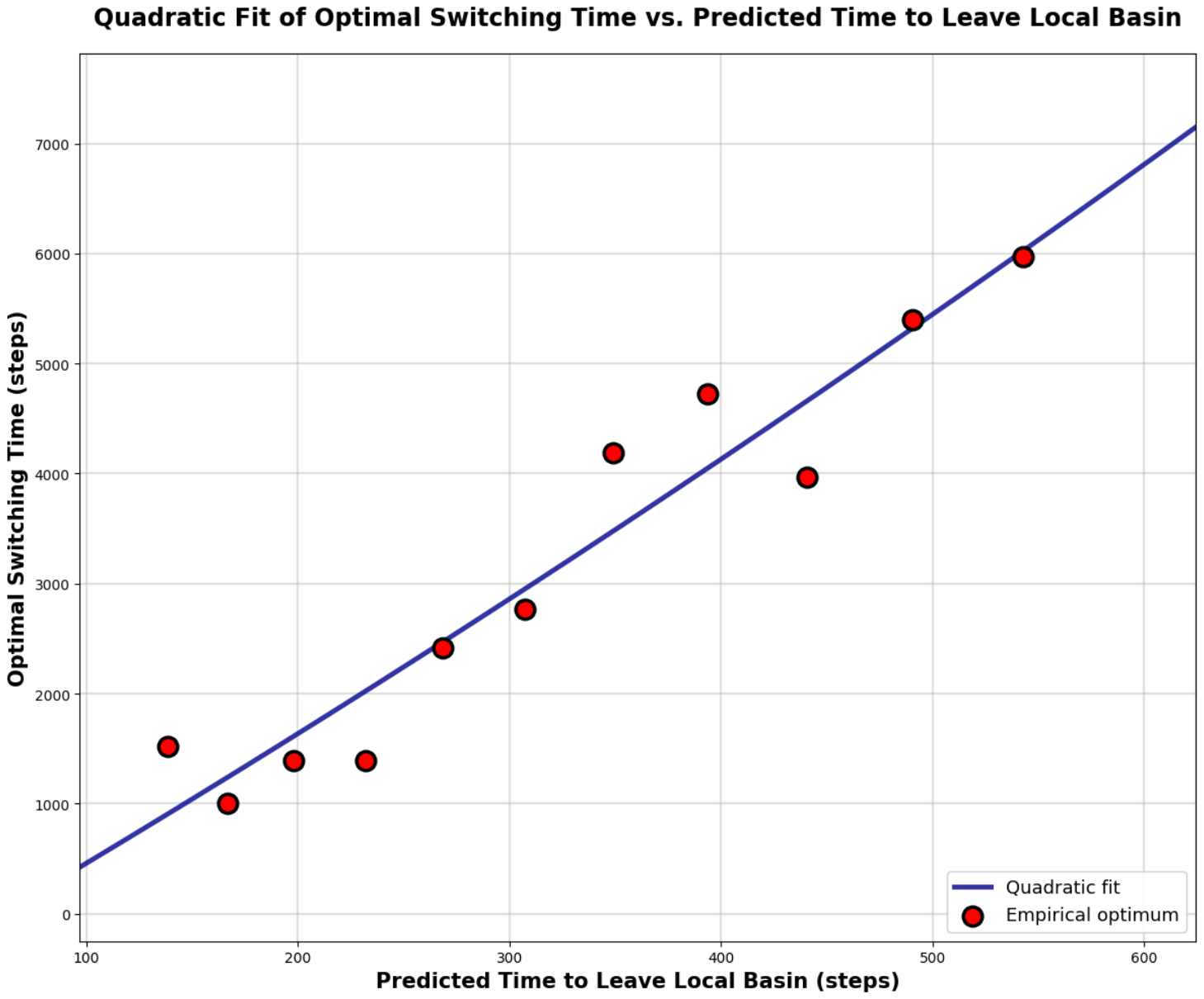}
  \caption{Plot of optimal switching time versus predicted time to leave local basin with a quadratic line of best fit}
  \label{fig:temperature_sweep_application}
\end{figure}

The empirically determined optimal switching time \(\tau\) is well approximated by a quadratic relation of the form
\(\tau = a_0\,\widehat{t}_{\mathrm{escape}}^{\,2} + a_1\,\widehat{t}_{\mathrm{escape}} + a_2\),
with fitted coefficients \(a_0 = 2.31\times10^{-3}\), \(a_1 = 11.09\), and \(a_2 = -675.93\), yielding a coefficient of determination \(R^{2}=0.924\).

This result indicates a superlinear dependence, well captured by a quadratic
model over the range considered, between the optimal switching time from
\(\mathrm{T}_{\mathrm{high}}\) to \(\mathrm{T}_{\mathrm{low}}\) and the predicted
time to leave the local basin. Consequently, the predicted basin escape
time—which can be computed inexpensively from the discrete analytical
expression—may be used to estimate an appropriate switching time without
requiring repeated simulation-based tuning.

\section{Discussion and Future Work}

This paper is a preliminary step toward a predictive finite-time theory of Simulated Annealing. In simplified one-dimensional landscapes, we derived closed-form expressions for expected basin escape and hitting times under constant temperature. Although these toy models do not capture realistic high-dimensional structure, they provide analytical transparency and help isolate how landscape geometry and temperature govern finite-time performance.

Natural extensions include analyzing time-varying temperature schedules (e.g., multi-stage policies informed by estimated escape times) and moving beyond one dimension. While exact formulas may not persist in higher-dimensional or nonlinear landscapes, the approach here may still yield useful approximations or bounds, especially for separable structure or dominant escape pathways. Another direction is to study sequences or networks of basins, which better reflect practical energy landscapes. Finally, these closed-form predictions suggest a practical use: reducing empirical hyperparameter tuning by providing analytical guidance for temperature and proposal-scale selection. Testing how much tuning effort can be saved in practice is an important next step.

\bibliographystyle{abbrv}
\bibliography{refs} 

\end{document}